

\vbadness=10000
%
%

\font\bf=cmbx10 scaled 1200

\font\it=cmti10 scaled 1200

\font\tenrm=cmr10 scaled 1200
\font\sevenrm=cmr7 scaled 1200
\font\fiverm=cmr5 scaled 1200
\font\teni=cmmi10 scaled 1200
\font\seveni=cmmi7 scaled 1200
\font\fivei=cmmi5 scaled 1200
\font\tensy=cmsy10 scaled 1200
\font\sevensy=cmsy7 scaled 1200
\font\fivesy=cmsy5 scaled 1200

\font\tenbf=cmbx10 scaled 1200
\font\sevenbf=cmbx7 scaled 1200
\font\fivebf=cmbx5 scaled 1200
\font\tensl=cmsl10 scaled 1200
\font\tentt=cmtt10 scaled 1200
\font\tenit=cmti10 scaled 1200
\catcode`\@=11
\textfont0=\tenrm \scriptfont0=\sevenrm \scriptscriptfont0=\fiverm
\def\rm{\fam\z@\tenrm}
\textfont1=\teni \scriptfont1=\seveni \scriptscriptfont1=\fivei
\def\mit{\fam\@ne} \def\oldstyle{\fam\@ne\teni}
\textfont2=\tensy \scriptfont2=\sevensy \scriptscriptfont2=\fivesy
\def\cal{\fam\tw@}
\textfont3=\tenex \scriptfont3=\tenex \scriptscriptfont3=\tenex
\newfam\itfam \def\it{\fam\itfam\tenit} 
\textfont\itfam=\tenit
\newfam\slfam  
\textfont\slfam=\tensl
\newfam\bffam \def\bf{\fam\bffam\tenbf} 
\textfont\bffam=\tenbf \scriptfont\bffam=\sevenbf
\scriptscriptfont\bffam=\fivebf
\newfam\ttfam  
\textfont\ttfam=\tentt
\catcode`\@=12
\rm


\abovedisplayskip=30pt plus 4pt minus 10pt
\abovedisplayshortskip=20pt plus 4pt
\belowdisplayskip=30pt plus 4pt minus 10pt
\belowdisplayshortskip=28pt plus 4pt minus 4pt

\hfuzz=10pt \overfullrule=0pt
\vsize 8.75in
\hsize 6.5in
\baselineskip=22pt

\parindent 20pt \parskip 6pt

 \mathcode`*="002A

\def\_{\vrule height 0.8pt depth 0pt width 1em}

\newbox\grsign \setbox\grsign=\hbox{$>$} \newdimen\grdimen \grdimen=\ht\grsign
\newbox\simlessbox \newbox\simgreatbox
\setbox\simgreatbox=\hbox{\raise.5ex\hbox{$>$}\llap
     {\lower.5ex\hbox{$\sim$}}}\ht1=\grdimen\dp1=0pt
\setbox\simlessbox=\hbox{\raise.5ex\hbox{$<$}\llap
     {\lower.5ex\hbox{$\sim$}}}\ht2=\grdimen\dp2=0pt

\def\dot#1{\vbox{\baselineskip=-1pt\lineskip=1pt
     \halign{\hfil ##\hfil\cr.\cr $#1$\cr}}}

\def\boxx{\mathop{\hbox to 0pt{$\sqcup$\hss}\hbox to
     0pt{$\sqcap$\hss}\phantom\nabla}}


\baselineskip=22pt



\centerline{\bf Beyond the Thin Lens Approximation}
\vskip0.25truein
\centerline{Ted Pyne and Mark Birkinshaw}
\centerline{Harvard-Smithsonian Center for Astrophysics}
\centerline{60 Garden St., Cambridge, MA 02138}
\vskip0.25truein
\centerline{Submitted to ApJ: 19 April 95}
\vskip0.5truein

\noindent
{\bf Abstract}

We obtain analytic formulae for the null geodesics of
Friedmann-Lema\^{\i}tre-Robertson-Walker
spacetimes with scalar perturbations in the longitudinal gauge.
{}From these we provide
a rigorous derivation of the cosmological lens equation, and
obtain an expression for the magnification
of a bundle of light rays
without restriction to static or thin lens
scenarios. We show how
the usual magnification matrix
naturally emerges in
the appropriate limits.

\noindent
{\bf Keywords:} Cosmology: Gravitational Lensing, Gravitation
\vskip0.5truein

\noindent
{\bf 1. Introduction}

The bending of light by a single symmetric gravitational
lens in Euclidean space is shown in Fig. 1. The symmetry
guarantees that the lines of sight from the observer, $o$,
to the lens, $l$, and to both the lensed and
unlensed images of the emitter, $e$, are coplanar,
and can be related by the angles $\alpha$, $\beta$, and $\theta$.
$D(o,l)$ and $D(o,e)$ are the distances from the
observer to the lens and emitter, respectively.
We assume that the deflection angle, $\alpha$, is small.
Then,
locally about the line of sight to the emitter's image,
we may approximate the two-spheres at distances
$D(o,l)$  and $D(o,e)$ from the observer as planes, called the
lens plane and source plane respectively.
$D(l,e)$ denotes the distance between these two planes.
The assumption of small deflection angle also allows us to relate the
lensing angles by $\beta =\theta -\alpha D(l,e)/D(o,e)$.
This is the simplest expression of the gravitational lens equation
(Refsdal 1964).

The generalization of this equation to more complicated
lens structures and non-Euclidean background spaces
proceeds by a number of steps. We continue to assume that the actual
path of the photon is well approximated by two segments joined
at a point of deflection, $p$, located near the lens.
A general lens is not
symmetric so that the angles $\alpha$,
$\beta$, and $\theta$ are not necessarily coplanar. To handle this,
consider a set of cartesian
axes with origin at the observer. Choose the $x$-axis to
coincide with the line-of-sight to the image. Let $\alpha^i$, where
$i$ runs over $\lbrace 2,3\rbrace$,
be the angle between the $x$-axis and the projection
into the $x^ix$-plane of the line-of-sight vector
from the deflection point, $p$, to the emitter. Similarly, let
$\beta^i$ be the angle between the
projection into the $x^ix$-plane of the line-of-sight
vector from the observer to the lens and the
projection into the $x^ix$-plane of the line-of-sight
vector from the observer to the unlensed image.
Also, let $\theta^i$ be the angle between the $x$-axis
and the projection into the
$x^ix$-plane of the line-of-sight vector from the
observer to the lens.  To allow for non-Euclidean spatial geometries
$D(o,e)$ is taken to be the angular-diameter
distance in the background geometry from the observer
to the intersection of the $x$-axis with the source plane,
and $D(l,e)$ is taken to be the angular-diameter distance in the
background geometry
between the deflection point, $p$, and the intersection
of the $x$-axis with the source plane, $p^{\prime}$.
Then, again assuming small deflection angle,

$$\beta^i =\theta^i -{D(l,e)\over D(o,e)}\alpha^i .         \eqno{(1)}$$

\noindent
This is the standard cosmological gravitational lens equation
(e.g. Schneider, Ehlers, and Falco 1993, Chapter 2).
An important quantity associated with this
formalism is the magnification matrix, $M^i{}_j=
\partial \beta^i /\partial\theta^j $, which
contains information on the deformation of ray bundles
connecting the observer and emitter.
For example, the inverse of the determinant of this
matrix is the magnification of an image
relative to an unlensed image.

The purpose of the current paper is to address
a number of subtle issues
that arise when we attempt to justify
mathematically the use of the
lens equation (1) for calculations in our Universe, although most workers
agree that the physical basis for its use in observed lens
systems is strong.
First, there is the question of the correct choice of distance factors.
There exists a large
literature addressing this question,
primarily concerned with the appropriateness
of the so-called Dyer-Roeder distances
(Dyer and Roeder 1972, 1973; Ehlers and Schneider 1986;
Futamase and Sasaki 1989; Watanabe and Tomita 1990; Watanabe, Sasaki,
and Tomita 1992; Sasaki 1993).
Our work suggests that within the framework of
cosmological perturbation theory, the natural
distance factors to use are those of the background.
Hence, the choice of distance factors
is equivalent to the choice of cosmological
model, in agreement with the recent results
of Sasaki (1993). The issue of the most appropriate choice of
cosmological model
must be addressed in its own right.

The second concern in
any mathematical investigation into the lens
equation is the accuracy of the approximation of
an actual photon path by
two geodesics of the background which join
at a point near the lens: the deflection point, $p$.
On physical grounds we expect this approximation
to be good for systems for which the photon-lens
interaction is localized: the thin-lens approximation.
One purpose of our present work is to quantify the error involved
in using two geodesics of the background, rather than
the actual path, in deriving the lens equation, (1).

Third, how are we to find the angles
appearing in the lens equation from physical data?
Generally, the $\alpha^i$ are taken to be
those calculated in
Einstein-de Sitter spacetime,
since the overall curvature of space should not be important
near $p$, where the light ray
interacts with the lensing object.
For static, thin lenses,
the deflection angle is written as a superposition of
point mass deflection angles
contributed by mass elements of the lens projected onto the
lens plane (Schneider et al. 1993).
For brevity we will term the resultant angle
the ``Einstein angle.''
Another purpose of the present work is to derive this result
from the full equations of light propagation under
an appropriate set of mathematical approximations.
Succinctly, the lens equation, for static, thin lenses,
effectively assumes that
the light path is described by the Jacobi equation of the
background spacetime subject to an impulsive
wavevector deflection at the lens
plane by an angle equal to the usual
Einstein bending angle. We wish to quantify the level of
approximation involved.

There have been two notable recent attempts to clarify
the validity of the cosmological
lens equation by deriving it
from the optical scalar equations
(Seitz, Schneider, and Ehlers 1994) and the Jacobi
equation (Sasaki 1993).
However, a crucial difference between these papers
and the present work is that they treat the path
of the light ray differently near to
and far from the lens.
It is precisely this
assumption that we must eliminate if we
hope to gain a more general lens equation able
to quantify the errors implicit in equation (1).

Our
approach is to investigate
the cosmological lens equation as it emerges from the geodesic
equation. In this our work is complementary to that of Seljak (1994)
and Kaiser (1992)
who have used an approach like this to investigate certain lensing
questions for a perturbed Einstein-de Sitter spacetime. Where our work
overlaps that of these authors we are in agreement.
We will show, however, that it is possible to handle the curved
Friedmann-Robertson-Walker (FRW)
spacetimes by analogous calculations, though of somewhat more
technical difficulty. With theorists beginning to take the idea of
spatially curved models more seriously (Kamionkowski
{\it et al.} 1993, 1994; Spergel {\it et al.}
1993), we
feel that this extension is of more than formal importance.
In this paper we make use of a
method for constructing null geodesics in
perturbed spacetimes introduced in Pyne and Birkinshaw (1993),
hereafter PB. The method is the analog for geodesic curves of
familiar perturbation techniques for differential equations.
The results presented here come from applying this method to
FRW spacetimes with scalar perturbations in the longitudinal gauge.
Our principal results are:

\item {(1)}
analytic formulae (equations (13) and (14)) for light
rays in the spacetime (5);

\item {(2)} a general expression for the magnification
undergone by a bundle of light rays
capable of handling non-static,
geometrically thick, lenses (equations (35), (36), and (38));

\item {(3)}
a demonstration that the usual lens
equation (1) and magnification
matrix are recovered in the appropriate approximations.

\noindent
To our knowledge, this is the first rigorous derivation
of the cosmological lens
equation for perturbed FRW spacetimes with spatial curvature.

The outline of this paper is as follows. In section 2 we review
the perturbative geodesic expansion introduced in PB. In section 3
we apply this formalism to the problem of constructing null geodesics
of FRW spacetimes with scalar perturbations. With the help of the
equation of geodesic deviation for the FRW background, the
solutions we find will enable us to understand the role of
the Einstein angle for light propagation in these spacetimes.
In section 4 we obtain
an expression for the magnification of a bundle of light rays in
such spacetimes without restriction to static, geometrically
thin perturbations. We then discuss the emergence of the
usual cosmological lens equation, (1), in the appropriate limits.
In section 5 we illustrate the
use of the magnification equation by solving it for a point
mass embedded in an Einstein-de Sitter spacetime.

\vskip0.5truein
\noindent
{\bf 2. The Perturbative Geodesic Expansion}

\nobreak
We work in geometrized units, $G=c=1$. We let Greek indices
$\mu ,\nu ,...$ run over $\lbrace 0,1,2,3\rbrace $ and Roman indices
$i,j,...$ run over $\lbrace 1,2,3 \rbrace$. The spacetime metric
is taken to have signature $+2$.
Our Riemann and Ricci tensor conventions are given by
$\left[ \Delta_{\alpha},\Delta_{\beta}\right] v^{\mu}=
R^{\mu}{}_{\nu\alpha\beta}v^{\nu}$ and $R_{\alpha\beta}=
R^{\mu}{}_{\alpha\mu\beta}$.

In PB we showed that given a
null geodesic of $g^{(0)}_{\mu\nu}$, $x^{(0)\mu}
(\lambda )$, with $\lambda$ an affine parameter,
we could construct a set of four functions, which we call
the separation, $x^{(1)\mu}(\lambda )$, transforming as a
vector under infinitesimal co-ordinate change, which were sufficient to
ensure that $x^{(0)\mu}(\lambda )+x^{(1)\mu}(\lambda )$ is an
affinely parametrized null geodesic of $g^{(0)}_{\mu\nu}+
h_{\mu\nu}$. In order to construct $x^{(1)\mu}(\lambda )$ we need
three quantities, two referring solely to the background
metric and one referring to the perturbation. First, we need the
parallel propagator (Synge, 1960) appropriate to $x^{(0)\mu}(\lambda )$
in $g^{(0)}_{\mu\nu}$ (in PB we called this the connector after
DeFelice and Clarke (1990, Section 2.3),
mostly to avoid confusion with the Jacobi
propagator below, but we prefer Synge's term). We remind the reader
of one crucial property of the parallel propagator: if $v^{\mu}
\left( \lambda_2\right)$ is a vector at $x^{(0)\mu}\left( \lambda_2
\right)$ and $v^{\nu}\left( \lambda_1\right)$ is its parallel
translate along $x^{(0)\mu}(\lambda )$ to the point $x^{(0)\mu}
\left( \lambda_1\right)$ then the parallel propagator $P\left(
\lambda_1 ,\lambda_2\right)^{\mu}{}_{\nu}$ obeys
$v^{\mu}\left( \lambda_1\right)=P\left( \lambda_1 ,\lambda_2\right)
^{\mu}{}_{\nu}v^{\nu}\left( \lambda_2\right)$. That is, it parallely
propagates vectors along $x^{(0)}(\lambda )$.

The second quantity we need, the Jacobi propagator,
also refers only to the background
spacetime. Introduced in PB,
it is an $8\times 8$ dimensional matrix which serves as
a Green's function for the Jacobi equation for $g^{(0)}_{\mu\nu}$
along $x^{{0}\mu}(\lambda )$. Its explicit construction makes use
of the parallel propagator, the curvature tensor of
$g^{(0)}_{\mu\nu}$, $R^{(0)\mu}{}_{\nu\rho\sigma}$, and the tangent
vector to $x^{(0)\mu}(\lambda )$, the wavevector, $k^{(0)\mu}(\lambda )
=dx^{(0)\mu}(\lambda )/d\lambda$. With eight dimensional systems the
usual tensor notation can be cumbersome so we will use the matrix
notation of PB. We let ${\cal R}(\lambda )^{\mu}{}_{\sigma}$ denote the
$4\times 4$ matrix $R^{(0)\mu}{}_{\nu\rho\sigma}
k^{(0)\nu}k^{(0)\rho}$ evaluated at $x^{(0)}(\lambda )$ and
write $P\left(\lambda_1 ,\lambda\right)^{\mu}{}_{\nu}
{\cal R}\left( \lambda\right)^{\nu}{}_{\rho}P\left( \lambda ,\lambda_1
\right)^{\rho}{}_{\sigma}$ as $P\left( \lambda_1 ,\lambda\right)
{\cal R}\left( \lambda\right)P\left( \lambda ,\lambda_1\right)$.
Then the Jacobi propagator $U\left( \lambda_1 ,\lambda_2\right)$
is given by

$$U\left( \lambda_1 ,\lambda_2\right)={\cal P}\exp \left(
\int_{\lambda_1}^{\lambda_2}\pmatrix{ 0 &1_d \cr P\left(
\lambda_1 ,\lambda\right){\cal R}\left( \lambda\right)P\left( \lambda ,
\lambda_1\right) & 0\cr}d\lambda \right)  .\eqno{(2)}$$

\noindent
Here $1_d$ is the $4\times 4$ identity matrix and $\cal P$ denotes the
path ordering symbol.

The final quantity we need encodes the effects of the perturbation
itself. It is a vector field defined along
$x^{(0)\mu}(\lambda )$. We denote the covariant derivative with
respect to $g^{(0)}_{\mu\nu}$ by a semi-colon. Then the perturbation
vector at affine parameter $\lambda$, $f^{\mu}(\lambda )$, is given
by

$$f^{\nu}={1\over 2}h_{\alpha\beta}{}^{;\nu}k^{(0)\alpha}k^{(0)\beta}
-h^{\nu}{}_{\alpha ;\beta}k^{(0)\alpha}k^{(0)\beta}  \eqno{(3)}$$

\noindent
evaluated at $x^{(0)\mu}(\lambda )$.

It was shown in PB that the force vector, Jacobi propagator,
and parallel propagator allow us to solve for
the separation, $x^{(1)\mu}(\lambda )$. In particular, we suppose that
we desire to construct a geodesic of $g^{(0)}_{\mu\nu}+h_{\mu\nu}$
of the form $x^{\mu}(\lambda )=x^{(0)\mu}(\lambda )+x^{(1)\mu}
(\lambda )$ and that we know the appropriate boundary conditions at
affine parameter $\lambda_1$. Then $x^{(1)\mu}$ is given at
arbitrary affine parameter, $\lambda_2$, by

$$\eqalign{ \pmatrix{ P\left( \lambda_1 ,\lambda_2 \right)x^{(1)}\left(
\lambda_2\right) \cr {d\over d\lambda_2}\left[ P\left( \lambda_1 ,
\lambda_2\right)x^{(1)}\left( \lambda_2\right) \right]}
 &=U\left( \lambda_2 ,\lambda_1\right)\pmatrix{ x^{(1)}\left(
\lambda_1 \right) \cr \left[ {d\over d\lambda}\left[ P\left( \lambda_1 ,
\lambda \right)x^{(1)}(\lambda )\right]\right]_{\lambda=\lambda_1} } \cr
 &\qquad +\int_{\lambda_1}^{\lambda_2}U\left( \lambda_2 ,\lambda \right)
\pmatrix{ 0 \cr P\left( \lambda_1 ,\lambda \right)f(\lambda )}
d\lambda } \eqno{(4)}$$

\noindent
where the integral in this equation is taken over the background
path, $x^{(0)\mu}(\lambda )$.

While the solution above looks complicated, in practice
background spacetimes are chosen specifically for their high degree
of symmetry and tractability and this often allows us to construct the
needed propagators explicitly. In these cases, equation (4) reduces
the work of finding null geodesics in the perturbed spacetime
to a simple problem of integration. We will see below that the crucial
FRW spacetimes belong to this class.

The consistency criteria for our solution are precisely those of
all perturbative type geodesic calculations;
extremely heuristically, the background and
the constructed geodesic should not be allowed to reach
regions where their spatial or temporal deviations are such that
the geodesics effectively feel different gravities
at equal affine parameter, either due to
the perturbation or the curvature of the background itself.
A simple {\it a priori} estimate for the domain of validity
was given in PB, but in practice, the consistency may usually
be checked easily after a solution is obtained.
\footnote{$^1$}{ We note that the condition $\epsilon^2\ll
\kappa$ given in PB is not, in fact, necessary. We thank Uros Seljak
for pointing this out to us.}

\vskip0.5truein
\noindent
{\bf 3. The Deflection Angle}

\nobreak
We will now employ the techniques reviewed above to investigate
the theory of gravitational lenses in cosmology.
Our starting point is a choice for the metric.
We choose to work
with FRW spacetimes with scalar perturbations
written in the longitudinal gauge,

$$d{\bar s}^2=a^2\left[-(1+2\phi )d\eta^2 + (1-2\phi )
\gamma^{-2}\left( dx^2+dy^2+dz^2 \right)\right]  \eqno{(5)}  	$$

\noindent
where
$\gamma =1+\kappa r^2/4$, $\kappa$ being the spatial curvature
parameter
($\pm 1$ or 0) and
$r^2=x^2+y^2+z^2$. Inhomogeneities are represented by the
quasi-Newtonian potential, $\phi$. For the order needed by us,
the expansion
factor, $a(\eta )$, is unperturbed from its Friedmann form
(Jacobs, Linder, and Wagoner 1993). We choose this form for the metric
for a number of reasons. The metric (5) is also used by
Seitz et al. (1994) and Sasaki (1993),
so that direct comparison of results is possible, and
recent work by Futamase (1989) and Jacobs et al.
(1993) has shown that
structure of galactic scale
and greater in our Universe can be
well modeled by metrics of this type.
While results obtained with the metric (5) are not appropriate
for lensing by gravitational waves or vector perturbations,
these cases are easily handled in an analogous manner.

Next we choose a particularly useful class of background light
rays with which to build our perturbed solutions; radial
null geodesics intersecting the observer at the spatial origin.
Because the Friedmann expansion, $a$, plays the role of a conformal
factor  it is simplest to work with the null geodesics of
$ds^2$, defined by $d{\bar s}^2=a^2ds^2$.
Light rays in these two metrics coincide and their
(affine) parameterizations are related by ${\bar
k}^{\mu}=a^{-2}k^{\mu}$.
With the observer located at the spatial origin,
the radial null geodesics of $ds^{(0)2}$,
(i.e. of that part of $ds^2$ independent of $\phi$),
may be written
$k^{(0)0}=1$ and $k^{(0)i}=
-\gamma e^i$, where $e^i$ are the direction cosines at the observer,
so that $\sum_{i=1}^{3}(e^i)^2=1$ (McVittie 1964). Note that we
have choosen our affine parameter to coincide with conformal time.
This is purely for convenience.
The explicit solutions for the comoving radius and for $\gamma$ along
such rays are given by

$$\eqalign{r (\lambda )    &=2\tan_{\kappa }\left(
{\lambda_o-\lambda \over 2}\right)  \cr
          \gamma (\lambda ) &=\sec_{\kappa }^2\left(
{\lambda_o-\lambda \over 2}\right) \cr} \eqno{(6)} $$

\noindent
where $\lambda_o$ is the affine
parameter at the observer. The subscript $\kappa$ on a
trigonometric function denotes a set of three functions:
for $\kappa =1$ the trigonometric function itself,
for $\kappa =-1$ the corresponding hyperbolic
function, and for $\kappa =0$ the first term
in the series expansion of the function.
The paths of the rays
are $x^{(0)0}=\lambda$, $x^{(0)i}=re^i$.

The equations of parallel transport along $x^{(0)\mu}(\lambda )$ are
easy to solve. Two vectors, $v^{\mu}_2$ and
$v^{\mu}_1$, at $x^{(0)\mu}\left( \lambda_2\right)$
and $x^{(0)\mu}\left( \lambda_1\right)$ respectively, are related
by parallel translation provided that $v^0_2=v^0_1$ and that
$v^i_2=\gamma \left( \lambda_2\right)v^i_1/\gamma\left( \lambda_1
\right)$. We can, thus, read off the parallel propagator for our
class of geodesics,

$$P\left( \lambda_2 ,\lambda_1\right)^{\mu}{}_{\nu}=
\pmatrix{ 1 &0_j \cr 0^i & {\gamma\left( \lambda_2 \right)\over
\gamma\left( \lambda_1 \right)}\delta^i_j \cr}  .\eqno{(7)}$$

Next we will obtain the Jacobi propagator. The Riemann tensor
of $ds^{(0)2}$ is non-vanishing only when all indices are spatial,
when

$$R^{(0)i}{}_{jkl}=-\kappa \left( g^{(0)i}{}_lg^{(0)}_{jk}
-g^{(0)i}{}_kg^{(0)}_{jl}\right)
\eqno{(8)}$$

\noindent
where $g^{(0)}_{\mu\nu}$ is the metric $ds^{(0)2}$.
As a result we can write $R^{(0)\mu}{}_{\nu\rho\sigma}k^{(0)\nu}
k^{(0)\rho}=-\kappa J^{\mu}{}_{\sigma}$ where

$$J^{\mu}{}_{\sigma}=\pmatrix{ 0 & 0_j \cr 0^i & \delta^i_j
-e^ie_j \cr} .\eqno{(9)}$$

\noindent
$J$ is idempotent, $J^2=J$. This allows us easily to sum the
series defining the Jacobi propagator, (2), in $4\times 4$ subblocks.
The result is, returning fully to our matrix notation,

$$\eqalign{ U\left( \lambda_2 ,\lambda_1 \right) &=J\otimes
\pmatrix{ \cos_{\kappa}\left( \lambda_2 -\lambda_1\right) &
\sin_{\kappa}\left( \lambda_2 -\lambda_1\right) \cr
-\kappa \sin_{\kappa}\left( \lambda_2 -\lambda_1\right) &
\cos_{\kappa}\left( \lambda_2 -\lambda_1\right) \cr}\cr
&\qquad +\left( 1_d -J\right)\otimes \pmatrix{ 1 & \left(
\lambda_2 -\lambda_1\right) \cr 0 & 1 \cr}\cr} .
\eqno{(10)}$$

\noindent
We note that $J$ may be interpreted as a projection
operator into the space transverse to the photon direction in the
comoving spatial hypersurfaces. Thus
the Jacobi propagator, $U$, has split into a
transverse rotation and a longitudinal shear.

Another crucial quantity we will need
before we can construct the perturbed
light rays is the force vector $f^{\mu}$. We note that the force vector
may be constructed not only from (3), but also from
the equivalent

$$ f^{\mu}=\Gamma^{(1)\mu}{}_{\alpha\beta}k^{(0)\alpha}
k^{(0)\beta} \eqno{(11)}$$

\noindent
where $\Gamma^{(1)\mu}{}_{\alpha\beta}$ denotes that part of the
Christoffel connection of $ds^2$ which is linear in either the
metric perturbation or its first partial derivatives. Direct
calculation yields

$$f^{\mu}=-2\pmatrix{ k^{(0)j}\phi_{,j} \cr \phi^{,i}
-k^{(0)i}k^{(0)\alpha}\phi_{,\alpha} \cr} \eqno{(12)}$$

\noindent
where a comma denotes an ordinary partial derivative.

It remains only to determine the appropriate initial conditions for
the separation before we can use (4) to gain the null geodesics of
$ds^2$. We will choose the perturbed geodesic and the background
geodesic to coincide at the observer, $x^{(1)\mu}\left(
\lambda_o\right)=0$. As discussed in PB, we can not take the
wavevectors of the two geodesics to coincide fully at the
observer because the geodesics must each be null in their respective,
different, metrics. We will choose, for convenience,
with $k^{(1)\mu}=dx^{(1)\mu}/d\lambda$, $k^{(1)i}\left(
\lambda_o\right) =0$. The constraint that $k^{(0)\mu}+
k^{(1)\mu}$ be null in $ds^2$ at the observer then tells us that
$k^{(1)0}\left( \lambda_o \right)=-2\phi_o$ where we denote the
value of $\phi$ at the observer by $\phi_o$.

It is now simply a matter of assembling the necessary pieces in (4)
to gain the separation at arbitrary affine parameter $\lambda_e$,
and thus the light rays of $ds^2$. Some straightforward labor yields

$$x^{(1)0}\left( \lambda_e \right) =-2\left( \lambda_e -\lambda_o\right)
\phi_o +2\int_{\lambda_o}^{\lambda_e}\,
d\lambda \left( \lambda -\lambda_e\right) k^{(0)m} \phi_{,m}\left( \lambda
\right) \eqno{(13)}$$

\noindent
and

$$\eqalign{x^{(1)i}\left( \lambda_e \right)
			&=-2k^{(0)i}\int_{\lambda_o}^{\lambda_e}\,
d\lambda \left( \lambda -\lambda_e\right)
{\partial \phi\over \partial \eta}(\lambda )	\cr
			&\qquad +2\gamma (\lambda_e )
\int_{\lambda_o}^{\lambda_e}\, d\lambda
\sin_{\kappa } \left( \lambda-\lambda_e\right) {1\over \gamma (\lambda )}
\left( \nabla_{\perp }\phi \right)^i (\lambda )\cr}
					\eqno{(14)}$$

\noindent
where $\nabla_{\perp }{}^i=g^{(0)mn}\left(
\delta^i_n -e_ne^i\right)\partial_m $ is the transverse
gradient operator. We see that the spatial
separation is naturally written as the
sum of a longitudinal and
a transverse term. We remind the reader
that the above integrals are taken over
the background geodesic $x^{(0)\mu}(\lambda )$.
Hence if these solutions are used for
geometrically thick lenses, the error terms will become important
at some finite distance along the geodesic.
In this case, it will in general
be necessary either to apply an iterative procedure,
incorporating a number of background paths, or to
appeal to statistical arguments to bound the errors. These
approaches are familiar from the usual multiple lens
plane theory (Schneider {\it et. al.} 1993; see also Seljak, 1994).
A crucial difference, however, between the use of multiple
background paths in our formalism and the multiple lens plane
method is that formulae (13) and (14), in principal,
allow the photon path to be approximated
to arbitrary accuracy by succesive modifications
of $x^{(0)\mu}(\lambda )$, in contrast with the
multiple lens plane theory where the continuum limit
is not compatible with the assumptions underlying
the theory.

It is not too difficult to check by straightforward calculation
that the geodesic defined by (13) and (14) is, in fact, null.
This is to be expected in light of the general theorem proved in PB
that our constructed geodesic preserves its null character. We
point out, however, that our appeal, in that paper, to the
co-ordinate invariance of scalar quantities in order to argue for
the vanishing of the term involving the partial derivative of
$g^{(0)}_{\mu\nu}$ is invalid. The theorem is, nevertheless, true.
\footnote{$^2$}{ One way to patch it up, for instance, is to note
that $g^{(0)}_{\mu\nu ;\rho}=0$ allows us to replace
the offending term by a sum of two terms linear in the
Christoffel symbols of the background, each of which vanishes
in the chosen co-ordinate system of the second half of the proof.}

We can gain more understanding of (13) and (14) by considering
their relation to the equation of geodesic deviation.
In Appendix A we show that the Jacobi
equation of $ds^{(0)2}$ subject to an
arbitrary impulsive wavevector perturbation $\delta k^{\mu}$ at some
affine parameter $u$, is solved by deviation
vector $\delta x^{\mu}$
with spatial components

$$\delta x^i(\lambda )=-{\gamma (\lambda )
\over \gamma (u)}\, \sin_{\kappa } (u-\lambda )\,
\delta k^i_{\perp}(u)-{\gamma (\lambda )\over \gamma (u)}\,
(u-\lambda )\, \delta k^i_{\parallel}(u)          \eqno{(15)}$$

\noindent
where $\delta k^i_{\perp}=\left( \delta^i_j-e^ie_j\right)\delta k^j$
is the impulse in the transverse direction,
and $\delta k^i_{\parallel}=
\delta k^i-\delta k^i_{\perp}$ is the longitudinal impulse.

A comparison of (14) and (15) leads to the interpretation of
the spatial components
of our solution for the separation, $x^{(1)i}$,
as the result of
a continuous sequence of impulsive perturbations

$$\delta k^i =-2\left( \nabla_{\perp }\phi \right)^i
 +2k^{(0)i}{\partial \phi \over \partial \eta} .\eqno{(16)}$$

\noindent
The form of the impulse can also be gained directly
from our equations by
differentiating the spatial separations with
respect to the affine parameter, and inserting a delta
function at $\lambda_l$ into the integrand which forces the
integrand to vanish except at the lens plane.
This gives

$$ k^{(1)i}\left(\lambda_l\right) =-2\left(
\nabla_{\perp}\phi\right)^i \left( \lambda_l\right)
+2k^{(0)i}\left( \lambda_l\right)
{\partial \phi \over \partial \eta}\left(
\lambda_l\right), 	\eqno{(17)}$$

\noindent
which
is exactly the impulse found above.

At this point we need only do a little work to
recover the Einstein angle from our equations.
Establish a set of cartesian axes at the observer
and choose the unperturbed wavevector $k^{(0)}=(1, -\gamma ,0,0)$.
Consider a static, localized perturbation in the
$xy$-plane.
Then the
angle represented by the impulse perturbation
found above (17), that is, the angle $k^{(0)i}\left(
\lambda_l\right)+k^{(1)i}\left(
\lambda_l\right)$ makes with $k^{(0)i}
\left( \lambda_l\right)$, is given by

$${\hat \alpha}^y={-2\left( \nabla_{\perp}\phi \right)^y\over
\gamma \left( \lambda_l\right)}=-2\gamma \left( \lambda_l
\right) \phi_{,y}.  \eqno{(18)}      $$

\noindent
The factor of $\gamma \left(\lambda_l \right)$
is present only because our co-ordinates are
scaled in an unusual
way at the lens plane. If we make the co-ordinate change $x^{i\prime}=
x^i/\gamma\left( \lambda_l \right)$
the metric on the lens plane becomes
Minkowskian and, locally near
the deflector,
$y'$ and $z'$ serve as normal co-ordinates on the lens plane.
In these co-ordinates
$\phi_{,y^{\prime}}$ on the lens plane takes on the usual Newtonian
form (with origin shifted from the lens, accounting for the
unusual minus sign).
Since our lensing angle $-2\gamma\left(
\lambda_l\right)\phi_{,y}=-2\phi_{,y^{\prime}}$
the integrated impulse
lensing angle for a localized perturbation
is exactly the Einstein deflection angle.
We emphasize that this is the first time this
has been shown rigorously for the curved FRW spacetimes.

For completeness we note that
the timelike component of the separation can also be analyzed
by comparison to the Jacobi equation.
The Jacobi equation of $ds^{(0)2}$ for
an impulse wavevector perturbation $\delta k^{\mu}$ at affine parameter
$u$ results in a timelike component of the deviation vector
$\delta x^0=-(u-\lambda )\delta k^0(u )$. Comparison
with our solution for the separation reveals that the time
delay undergone by the light ray relative to the fiducial background
ray may be considered to result from a sequence of impulses
$\delta k^0=-2k^{(0)m}\phi_{,m}$, in addition to a boundary term.

\vskip0.5truein
\noindent
{\bf 4. The Magnification}

\nobreak
We now want to examine the magnification
undergone by a bundle of light rays.
We define this after Schneider et al. (1993)
in the following way.
Suppose a source of given physical size at some redshift
is observed to subtend solid angle $d\Omega$. An identical
source observed at identical
redshift placed in an FRW spacetime would
subtend solid angle $d\Omega^{(0)}$.
The magnification $M$ is defined to
be $d\Omega / d\Omega^{(0)}$.

To gain the magnification we will
construct an infinitesimal bundle of light rays in $d{\bar s}^2$
which emanate from a source and converge at an observer,
located at the spatial origin of co-ordinates, by varying
the direction cosines of the background ray,
$e^i$, in equations (13) and (14).
We will determine the solid angle, $d\Omega $,
subtended by the rays in the
rest frame of an observer with four-velocity $u_o^{\mu}=\left( 1/
a_o\right)\left( 1-\phi_o ,v^i_o\right)$.
We will then ask what local area
transverse to its direction of propagation
the bundle sweeps out in a frame with
four velocity $u_e^{\mu}=\left( 1/a_e\right)\left( 1-\phi_e ,v^i_e
\right)$,
$dA$, at a given redshift, $z$. We regard the peculiar velocities
$v^i_o$ and $v^i_e$ as first order quantities so that $u^{\mu}_o$
and $u^{\mu}_e$ are properly normalized to first order in $d{\bar s}^2$.
We suppose a source with four-velocity
$u_e^{\mu}$ intersects our bundle,
that its redshift is $z$, that its physical size is
$dA$, and that its shape such that it exactly fills the beam; that is,
our bundle is the light of a physical source. In this way
we will gain a relationship between
$d\Omega$ and the redshift and proper size
of the source and the four-velocity of the observer.
A similar relationship is easy to
derive for an identical source in the background spacetime.
Comparison of the two expressions will yield $M$. Figure 2
illustrates the constructions of this section.

We begin by choosing a set of null geodesics
of the background with which
to construct our congruence. We use the two-parameter family of
curves given by $x^{(0)\mu}=\left( \lambda ,re^i\right)$ with
$r$ as in (6),
$e^i=(1, d\sin\theta ,d\cos\theta )$, $d\in
(0,\epsilon )$ with $\epsilon$ infinitesimal, and $\theta \in (0,
2\pi ]$. We work to first order in
$\epsilon$. To this order, this set of rays defines a null congruence
of $ds^{(0)2}$. To each of the rays
of this congruence is associated
a null ray of $ds^2$, and hence of $d{\bar s}^2$, by (13) and (14) above,
$x^{\mu}\left( \lambda ;d,\theta \right)$. In the rest
space of the observer, these rays define
a cone. To see this, we note that the conical shape is
clear for a comoving observer. For an observer with some peculiar
velocity the assertion then follows from a result of Sachs (1961) on
the geometry of null rays.

Next we will determine the two-dimensional projected
area of our bundle at a given redshift.
In Appendix B it is proven that, to first order in the perturbation,
$w^{\mu}=x^{\mu}( d=\epsilon ,\theta )-x^{\mu}( d=0 )$ is
a one-parameter family of geodesic deviation vectors of $d{\bar s}^2$
along the central geodesic of the bundle, $x^{\mu}( d=0 )$.
{}From this point on, any wavevector or path pertaining either to
the perturbed or unperturbed spacetimes not written with
an explicit $d$-argument is intended to refer to the appropriate
central geodesic, $d=0$. Taylor expansion writes

$$w^{\mu}=\left( {\partial x^{(0)\mu} \over \partial e^j}+
{\partial x^{(1)\mu}\over \partial e^j}\right)_{e^i=
(1,0,0)}\left( \epsilon\sin\theta\delta^j_2 +\epsilon\cos\theta
\delta^j_3\right)  ,\eqno{(19)}$$

\noindent
which we can write as

$$w^{\mu}=w^{\mu}_{(2)}\epsilon\sin\theta +w^{\mu}_{(3)}\epsilon
\cos\theta ,\eqno{(20)}$$

\noindent
with

$$w^{\mu}_{(j)}={\partial x^{\mu}\over \partial e^j} \Bigl|_{e^i=
(1,0,0)} \eqno{(21)} $$

\noindent
for $j=2,3$.
It will also be useful to define, with $j=2,3$,

$$w^{(0)\mu}_{(j)}={\partial x^{(0)\mu}
\over \partial e^j }\Bigl|_{e^i=(1,0,0)}=r\delta^{\mu}_j ,\eqno{(22)}$$

\noindent
and

$$w^{(1)\mu}_{(j)}={\partial x^{(1)\mu} \over \partial e^j}
\Bigl|_{e^i=(1,0,0)} ,\eqno{(23)} $$

\noindent
so that $w^{\mu}_{(j)}=w^{(0)\mu}_{(j)}+w^{(1)\mu}_{(j)}$.

Suppose now that the central geodesic of our bundle intersects the
worldline of our hypothetical source
at the point $e$.
The projection of $w^{\mu}$ into the two-dimensional subspace
orthogonal to both $u_e^{\mu}$ and to the projection of
${\bar k}^{\mu}$ into
this subspace, i.e. to the direction of photon propagation, defines an
ellipse. We can determine the characteristics of this ellipse
explicitly. The relevant projection operator is given by
(Kristian and Sachs 1965)

$$H^{\mu}{}_{\nu}=\delta^{\mu}_{\nu}-{{\bar k}^{\mu}{\bar k}_{\nu}\over
\left( u_e\cdot {\bar k}\right)^2 }
-{u_e^{\mu}{\bar k}_{\nu}\over  u_e\cdot {\bar k}}
-{{\bar k}^{\mu}u_{e\nu}\over u_e\cdot {\bar k}} \eqno{(24)}$$

\noindent
where we denote the $d{\bar s}^2$ inner product by a dot, e.g.
$ u_e\cdot {\bar k}=u_e^{\alpha}{\bar g}_{\alpha
\beta}{\bar k}^{\beta}$. We will
use $\perp$ to denote the result of acting on a given vector with
$H^{\mu}{}_{\nu}$, e.g. $w_{\perp}^{\mu}=H^{\mu}_{\nu}w^{\nu}$.

Extremizing $ w_{\perp}\cdot w_{\perp} $ with respect to
$\theta$ we find
the major and minor axes of the ellipse occur for

$$\theta ={1\over 2}\tan^{-1}\left( {-2 w_{\perp (2)}\cdot
w_{\perp (3)}\over \left[  w_{\perp (2)}\cdot w_{\perp (2)}
 - w_{\perp (3)}\cdot w_{\perp (3)} \right] }
\right) +{n\pi \over 2} \eqno{(25)}$$

\noindent
with $n=0,1$. In (25)
the inner products are evaluated at $e$.
Denoting the argument of the inverse tangent by $\chi$,
the squared lengths of the two axes are given by, for
$\theta =(1/2)\tan^{-1}\chi$

$$ \eqalign{  w_{\perp}\cdot w_{\perp} &=
{\epsilon^2\over 2\sqrt{1+\chi^2}}
\Bigl[ \left( w_{\perp (2)}\cdot w_{\perp (2)}+
w_{\perp (3)}\cdot w_{\perp (3)}\right)\sqrt{1+\chi^2}\cr
&\qquad -\left( w_{\perp (2)}\cdot w_{\perp (2)}-
w_{\perp (3)}\cdot w_{\perp (3)}-2\chi
w_{\perp (2)}\cdot w_{\perp (3)}\right) \Bigr] \cr}\eqno{(26)}$$

\noindent
and for $\theta =-\pi /2+(1/2)\tan^{-1}\chi$

$$ \eqalign{  w_{\perp}\cdot w_{\perp} &=
{\epsilon^2\over 2\sqrt{1+\chi^2}}
\Bigl[ \left( w_{\perp (2)}\cdot w_{\perp (2)}+
w_{\perp (3)}\cdot w_{\perp (3)}\right)\sqrt{1+\chi^2}\cr
&\qquad +\left( w_{\perp (2)}\cdot w_{\perp (2)}-
w_{\perp (3)}\cdot w_{\perp (3)}-2\chi
w_{\perp (2)}\cdot w_{\perp (3)}\right) \Bigr] \cr}\eqno{(27)}$$

\noindent
which lead to the area of the ellipse, $dA$, being

$$ dA =\epsilon^2\pi\left[
\left( w_{\perp (2)}\cdot w_{\perp (2)}\right)
\left( w_{\perp (3)}\cdot w_{\perp (3)}\right) -
\left( w_{\perp (2)}\cdot w_{\perp (3)}\right)^2 \right]^{1/2}
.\eqno{(28)}$$

At this point the underlying structure of our calculation is
becoming clear. We are involved in a realization
of standard ideas in
the geometry of linear maps of the plane: by Jacobi
propagation followed by projection we have linearly mapped
the two plane spanned by $w^{\mu}_{(j)}$, $j=2,3$, to the transverse
two-plane at the source. Much of this section is easier to
understand with this in mind.

We continue by breaking the projection operator up into
zeroth and first order parts in the perturbation $H^{\mu}{}_{\nu}=
H^{(0)\mu}{}_{\nu}+H^{(1)\mu}{}_{\nu}$. It is then possible to show
that to first order $H^{\mu}{}_{\nu}w^{\nu}H_{\mu\alpha}w^{\alpha}=
H^{(0)\mu}{}_{\nu}w^{\nu}H^{(0)}_{\mu\alpha}w^{\alpha}$.
This relationship has a simple geometrical interpretation. By a theorem
of Sachs (1961), the four-velocity of the emitter has no effect on
the size and shape of our ellipse so we take the emitter to have
$v^i_e=0$. In this case $H$ and $H^{(0)}$ project onto two planes
inclined with respect to each other by a small angle, the lens
angle at the emitter. But the projected areas in two such planes
differ only by a factor of $\cos\alpha$, $\alpha$ the angle of
inclination. Since $\alpha$ is first order, the areas are equal to
first order. Therefore in the expression for the area, (28) above,
we are able to replace all $H$-projected quantities by
$H^{(0)}$-projected quantities.

Next we use the explicit form of the projector to find
$H^{(0)\mu}{}_{\nu}w^{\nu}=(0,0,w^y,w^z )$. Substituting back into (28)
then yields

$$\eqalign{ dA &=\epsilon^2\pi
{a_e^2 \over \gamma_e^2 }\left(1-2\phi_e\right)
{\rm Det}\pmatrix{ w^y_{(2)} & w^y_{(3)} \cr w^z_{(2)} &
w^z_{(3)} \cr} \cr
 &= \epsilon^2\pi {a^2_er^2_e\over \gamma^2_e}\left(
1-2\phi_e\right){\rm Det} M^i{}_j \cr} \eqno{(29)}$$

\noindent
where the magnification matrix, $M^i{}_j$ is given by

$$M^i{}_j=1_d+{1\over r}w^{(1)i}_{(j)} ,
\eqno{(30)}$$

\noindent
with $i,j\in \lbrace 2,3\rbrace $ and $1_d$ the $2\times 2$ identity
matrix. In (29) we have written $r_e$ for $r\left( \lambda_e\right)$
and $\gamma_e$ for $\gamma \left( \lambda_e\right)$.
$M^i{}_j$ is a function of the affine parameter and in both
the above equations is evaluated at $\lambda =\lambda_e$, corresponding
to the point $e$ along the central perturbed geodesic.
We may recognize the first of the equalities in (29) as the
transformation of area law for infinitesimal linear mappings of the
plane once we recognize the factors in front of the determinant
(aside from the factor of $\pi$)
as corresponding to the metric factors in the induced area form
for our image two-plane. Equivalently, this is the induced
area two-form on the $yz-$plane acting on $ H^{(0)}w_{(2)}$ and
$H^{(0)}w_{(3)}$ (again up to the factor of $\pi$).

The above expression (30) for the magnification matrix is one of the
key results of this section. We recall from (23)
that for $i,j=2,3$, $w^{(1)i}_{(j)}$
is precisely the variation of the transverse separation with
the transverse direction cosine. The factor of $r^{-1}$ in (30)
turns the transverse separation into the transverse angular deflection.
Thus, we are begining to see the usual structure of the
magnification matrix emerge. We notice, however, from (14), that while
the time variation of the potential does not contribute to the
transverse deflection of a single ray it will contribute to the
gradient in (23) (this will become clear below), so that
non-static potentials contribute to the magnification
matrix. We note that (30) may be used for vector and tensor
perturbations in addition to the scalar ones considered here,
after a change in the solution for the separation,
hence $w^{(1)i}_{(j)}$, which arises from a change in
the specific form of the force vector, $f^{\mu}$.

We can use (29) to find $d\Omega $, the solid angle in the rest frame
of the observer defined by our bundle.
To this end, let $\lambda_e=\lambda_o+\Delta\lambda$, $\Delta\lambda$
small. We will use (29) to express the projected area of our bundle
at $\lambda_e$ to order $\left( \Delta\lambda \right)^2$. Noting that
$\gamma\sin_{\kappa}\left( \lambda_o -\lambda\right)=
r$, we have

$$dA\left( \lambda_o+\Delta\lambda \right)=\pi\epsilon^2
a_o^2\left( \Delta\lambda \right)^2\left( 1-2\phi_o\right) \eqno{(31)}$$

\noindent
The proper spatial distance in the rest frame of $u_o^{\mu}$ in
the metric $d{\bar s}^2$ corresponding to an affine distance
$\Delta \omega$ along ${\bar k}^{(0)\mu}$, $\Delta L$,
is given by (Ellis, 1971) $\Delta L =\vert \Delta \omega\vert
\left( u\cdot {\bar k}\right)_o$. Since
$\Delta\omega =a_o^2\Delta\lambda $ we can write the
projected area of the bundle a unit spatial distance away from
the observer as

$$\eqalign{ d\Omega &=dA\left( \Delta L=1\right) \cr
 &=\pi\epsilon^2 a^{-2}_o{\left( 1-2\phi_o\right)\over
\left( u_o\cdot {\bar k}_o\right)^2 }\cr
 &=\pi\epsilon^2  \left( 1+2v^i_ok^{(0)}_i\left(
\lambda_o\right)\right) \cr}\eqno{(32)}$$

\noindent
where in the last equality we have used $u^{\mu}_o=
\left( 1/a_o\right) \left(
1-\phi_o ,v^i_o\right)$ and ${\bar k}^{\mu}_o=
a_o^{-2}k^{(0)\mu}_o+a^{-2}_o k^{(1)\mu}_o=a_o^{-2}
\left( 1-2\phi_o ,-1 ,0,0)\right)$. In fact (32) is simply
the Lorentz transformation law for solid angle to linear order in
$v^i_o$. Were the observer comoving, the solid angle
would be given by $\pi\epsilon^2$.
The expression (32) comes from transforming to a frame in which the
observer moves with velocity $v^i_o$.

We can use this expression to replace $\epsilon^2\pi$ in (29), yielding

$$dA=d\Omega\left( 1-2v^i_ok^{(0)}_i\left(
\lambda_o\right)\right){a^2_er^2_e\over \gamma^2_e}\left(
1-2\phi_e\right){\rm Det} M^i{}_j\left( \lambda_e\right)  \eqno{(33)}$$

\noindent
This is the desired expression
relating the proper area of our emitter and
the solid angle that it is observed to subtend given its affine
distance, $\lambda_e$ (which we will soon trade in for its redshift).
The equivalent
expression for a congruence of the background spacetime can be
gained from (33) by taking the perturbed quantities to vanish.
This gives

$$dA^{(0)}=d\Omega^{(0)}{a_q^2r^2_q\over \gamma^2_q} \eqno{(34)}$$

\noindent
where $q$ is some point along the central background
geodesic, $x^{(0)\mu}=( \lambda , r,0,0)$ (see Fig. 2).

To proceed we must
ensure that the points $e$ and $q$ correspond to the same
numerical source redshifts in their respective spacetimes.
Let the affine parameter corresponding to the point
$q$ be given by $\lambda_q$.
We need to impose $1+z^{(0)}\left( \lambda_q\right) =
1+z \left( \lambda_e \right)$, where $z^{(0)}$ refers to the
redshift to the source in the background along the central
background geodesic and $z$ refers to the redshift of the
source in the perturbed spacetime along the central perturbed
geodesic.
The redshift in the perturbed spacetime is given by
$1+z =\left( u_e\cdot {\bar k}\right)\left( \lambda_e \right)
/\left( u_o\cdot {\bar k}\right)\left( \lambda_o \right)$.
$1+z^{(0)}$ may be found
from this expression simply by forcing the perturbed quantities
to vanish. If we write $\lambda_e =\lambda_q +\delta\lambda_q$ with
$\delta\lambda_q$ a first order function of $\lambda_q$, the equal
redshift constraint is solved by

 $$\delta\lambda_q={a_q\over {\dot a}_q}\left(
v^i_ok^{(0)}_i\left( \lambda_o\right)
 -v^i_ek^{(0)}_i\left( \lambda_q \right)
+\left[ \phi
-{\dot a}a^{-1}x^{(1)0}
+k^{(1)0}\right]^q_o \right)  . \eqno{(35)}$$

\noindent
Here a subscript $q$ denotes evaluation at the point $q$,
an overdot denotes an exact conformal time derivative along
the comoving worldlines,
$\left[ f\right]^q_o=f_q
-f_o$, and the separation and its affine
derivative are solved for along $k^{(0)\mu}$, which
intersects both $o$ and $q$ by construction. We have also noted
that, to first order, $\phi_e=\phi_q$. We have not used the
analogous formula for the emitter's peculiar velocity simply to
emphasize that the emitter can be moving, in principle, under
many different types of non-gravitational forces and so the notion
of a smoothly varying peculiar velocity field may be inappropriate.
\footnote{$^3$}{We have, however, considered $v_e$ and $v_o$ to contribute,
in a numerical sense, only at first order. For instance, the source
and emitter four-velocities are properly normalized only if $v_{o(e)}$
are considered to be first order quantities. It is not
difficult to relax this assumption.}

Recalling the definition of the magnification,
we now set $dA^{(0)}$ in (34) equal to $dA$ in (33) and
expand $a_e$, $r_e$, and $\gamma_e$
about their values at $q$. We also use the equivalence
of $M^i{}_j\left( \lambda_e\right)$ and $M^i{}_j\left( \lambda_q\right)$
to first order.
The end result is the following formula for the magnification
of a source observed at redshift $z^{(0)}\left( \lambda_q \right)$;

$$ \eqalign{ M	&={1\over \left({\rm Det}M^i{}_j \right)
\left( \lambda_q\right) }\Bigl[ 1-2\phi_o+2
 v^i_ek^{(0)}_i\left( \lambda_q \right) -2k^{(1)0}\left( \lambda_q\right)
\cr &\qquad +2\cot_{\kappa} \left(\lambda_o-\lambda_q\right)
\delta\lambda_q+
\kappa\sin_{\kappa }\left(
\lambda_o-\lambda_q\right)e_ix^{(1)i}\left( \lambda_q \right)
\Bigr] .\cr} \eqno{(36)}$$

\noindent

We emphasize that the source redshift in the actual perturbed
spacetime is given numerically by $z^{(0)}\left( \lambda_q \right)$.
We have chosen to express $\lambda_e$ in terms of $\lambda_q$
rather than the other way round because this
choice makes it simple to take the physical
redshift as the independent variable in the
magnification formula, (36), rather than $\lambda_q$.

An explicit formula for the magnification matrix may be found
directly from (30), (23) and (14).
We  need to take a partial derivative of (14)
with respect to $e^i$, $i=2,3$, while keeping $\lambda$ fixed.
The only subtlety arises from the implicit dependence of $\phi$
and its spacetime derivatives on $e^i$ which arises because
of the need to evaluate these quantities on a particular unperturbed
geodesic. This means

$$\eqalign{ {\partial \phi \over \partial e^j} &=
{\partial \phi \over \partial
x^{\alpha} }{\partial x^{(0)\alpha}\over \partial e^j} \cr
 &={\partial \phi \over \partial
x^{\alpha} }w^{(0)\alpha}{}_{(j)} \cr
 &=r{\partial \phi \over \partial
x^j }. \cr} \eqno{(37)}$$

\noindent
In principle $\gamma$, and hence $g^{(0)}_{\mu\nu}$, have a
similar implicit dependence on $e^i$, but in fact this dependence
is vanishing. This is easy to understand once it is realized that
varying $e^i$ for $i=2,3$ with fixed $\lambda$ amounts to variation
tangent to the two-sphere at fixed co-ordinate radius $r(\lambda )$.
Thus $r$, and so $\gamma$ and $g^{(0)}_{\mu\nu}$, are
stationary. Keeping the above in mind, we find an explicit
formula for the magnification matrix without restriction
to either static or thin perturbations (i.e. lenses). With
$i,j=2,3$,

$$\eqalign{ M^i{}_j=\delta^i_j  &
+{2\over \sin_{\kappa}\left( \lambda_o-\lambda_q\right)}
\delta^i_j\int_{\lambda_o}^{\lambda_q}d\lambda
\, \left(\lambda -\lambda _q\right)
{\partial \phi
\over \partial\eta}( \lambda ) \cr
                        &-{2\over \sin_{\kappa}
\left( \lambda_o-\lambda_q\right)}
\int_{\lambda_o}^{\lambda_q}d\lambda
\, \sin_{\kappa}\left( \lambda -\lambda_q\right)\gamma ( \lambda ) \cr
	&\times \left[ \delta^i_j{\partial \phi\over \partial x}( \lambda )
-r( \lambda )\gamma^{-2}( \lambda )g^{(0)ik}( \lambda )
\phi_{,kj}( \lambda )\right]. \cr }\eqno{(38)}$$

While (38) is computationally quite useful, its relationship to the
usual form for the magnification matrix is easier to see after
rewriting. Introducing the angular diameter distance of $d{\bar s}^{(0)}$,
${\bar D}\left( \lambda_1 ,\lambda_2 \right) =a\left( \lambda_2 \right)
\sin_{\kappa} \left( \lambda_1 -\lambda_2 \right) $,
(see Appendix A), and using the
two-dimensional projected angle of section 1 above,

$${\hat \alpha}^i=-2{\left( \nabla_{\perp}
\phi\right)^i\over \gamma }    , \eqno{(39)}$$

\noindent
we can write

$$\eqalign{ M^i{}_j=\delta^i_j	&+2{a\left(
\lambda_q \right) \over {\bar D}\left( \lambda_o,\lambda_q\right)}
\delta^i_j\int_{\lambda_o}^{\lambda_q}d\lambda \,
\left( \lambda -\lambda_q\right){\partial \phi\over \partial \eta}
(\lambda ) \cr
	&-{1\over {\bar D}\left( \lambda_o,\lambda_q\right)}
\int_{\lambda_o}^{\lambda_q}d\lambda \,
{\bar D}\left( \lambda ,\lambda_q\right)
{\partial {\hat \alpha}^i\over \partial e^j}
(\lambda ). \cr}   \eqno{(40)}$$

\noindent
Finally, we note that to zeroth order
$\partial /\partial e^i=\partial /\partial \theta^i$ with
$\theta^i$ the vectorial angle of (1). This allows us to make the
replacement

$${\partial {\hat \alpha}^i \over \partial e^j}\rightarrow {\partial
{\hat \alpha}^i\over \partial \theta^j} ,\eqno{(41)}$$

\noindent
in (40). The result is the most elegant form of our equation. It is
also the easiest form in which to recover the thin
lens limit. For this, suppose the potential is that appropriate
to a static, geometrically thin lens. Then the
first term in (40) vanishes and we can approximate the angular
diameter factor as constant over the region for which the
potential is important. Finally, we replace $\lambda_q$ with
$\lambda_e$ since, to first order, they are equivalent in $M^i{}_j$.
The result is

$$M^i{}_j\approx \delta^i_j-{{\bar D}
\left( \lambda_l,\lambda_e\right)\over {\bar D}\left( \lambda_o ,
\lambda_e \right) }
{\partial \over \partial\theta^j}
\int_{\lambda_o}^{\lambda_e}d\lambda \,
{\hat \alpha}^i     .\eqno{(42)}$$

\noindent
We have already seen that the integral of
${\hat \alpha}^i$ over the background path
produces the Einstein deflection angle.
As a result we conclude that in this limit our equation
has reproduced the usual magnification matrix defined
by the $\theta$-gradient of equation (1).

\vskip0.5truein
\noindent
{\bf 5. The Thin Lens in Einstein-de Sitter Space}

\nobreak
As an explicit illustration of the ideas above we will solve for the
magnification matrix appropriate to a point mass perturbation of
Einstein-de Sitter spacetime. We will find that the usual
expression of the magnification matrix is correct provided
that the impact parameter is much the smallest lengthscale in
the problem (save, of course, for the Schwarzschild radius of the
point mass). Certainly the calculation below is the hard way to
produce this result. Nevertheless, for the first time the
standard result is obtained along with correction terms
arising from the time variation of the perturbation and
the difference between the actual path and its piecewise
geodesic approximation.

Our starting point is the potential approximation to general
relativity (Martinez-Gonzalez, Sanz, and Silk 1990), which writes
the perturbation to the flat FRW metric in (5) appropriate to
a comoving point particle of mass $m$ located at
$\left( x_l ,y_l ,z_l\right)$ on the spatial hypersurfaces of
constant conformal time as

$$\phi \left( \eta ,{\vec x }\right) =
{-m\over a(\eta ) \sqrt{ \left( x-x_l\right)^2 +\left( y-y_l\right)^2
+\left( z-z_l\right)^2 }} .\eqno{(43)}$$

\noindent
It is worth noting that this perturbation is not static. Its
time dependence, however, is simply that of the background
spacetime, that is, it is set by the cosmology.
We certainly do not expect time variation of this magnitude
to affect photons streaming past the lens. We will see this
prejudice borne out in the calculations below. Nevertheless, this
emphasizes that more complicated details will
emerge in the rigorous picture
of lensing than in the usual models.

We will suppose that, other than the point mass, the observer, and
the lensed source, the spacetime
is filled with dust so that we can take $a=a_o\eta^2 /\eta_o^2$
with the subscript $o$ denoting evaluation at the observer and
$a_o$ constant with the dimension of length
(McVittie 1964). It is useful to
keep in mind that with our conventions the only dimensionful
numbers in this problem are $a_o$ and $m$ and both have
dimensions of length.

To avoid unnecessary complications we will take both the source
and the emitter to be comoving, as would be true, for instance,
if both were far from the point mass. We suppose the observer
to be at the spatial origin of co-ordinates with
the source image toward the positive $x$ direction. The light from
the source does not travel exactly down the $x$-axis because it
is bent by the action of the lens. We could calculate the
bending by constructing the path using (13) and (14) above. Instead
we will use (38) to gain the magnification matrix
directly. The background path appropriate to the situation
is given by $x^{(0)\mu}(\lambda )=\left( \lambda ,\lambda_o -\lambda ,
0,0\right)$ with $\lambda_o$ the affine parameter at the observer.
We denote by $\lambda_l$ that value of the affine
parameter at which $x^{(0)\mu}$ intersects the plane $x=x_l$, i.e.
$x_l=\lambda_o -\lambda_l$. For the usual lensing scenarios,
this point of intersection is very nearly the point
of closest approach of the photon to the lens.

Given our geometry, the physical linear size of the impact
parameter, $b$, is given by

$$b=a\left( \lambda_l\right) \sqrt{ y_l^2+z_l^2}\eqno{(44)}$$

\noindent
so that the angle between the lens and the image of the emitter,
$\theta$, is given by

$$\theta={ a\left( \lambda_l\right) \sqrt{ y_l^2+z_l^2}\over
{\bar D}\left( \lambda_o ,\lambda_l\right) }\eqno{(45)}$$

\noindent
It will be convenient to define an unbarred angular-diameter
distance symbol, $D_{12}$, by $D_{12}={\bar D}\left( \lambda_1 ,
\lambda_2\right) /a\left( \lambda_2 \right)=\left( \lambda_1 -
\lambda_2\right)$ so that

$$\theta ={\sqrt{ y_l^2+z_l^2}\over D_{ol} } \eqno{(46)}$$

\noindent
Also, in terms of the vectorial decomposition of $\theta$ introduced
in section 1, we have $x^i_l=\theta^i D_{ol}$, $i=2,3$. The
physical situation described by the above geometry
is illustrated in Fig. 3.

We start the calculation by noting that for an Einstein-de Sitter
background we may combine the second and third terms in (38) to give

$$\eqalign{M^i{}_j=\delta^i_j &
+{2\over D_{oq}}\delta^i_j \int^{\lambda_q}_{\lambda_o}
\, d\lambda \left( \lambda -\lambda_q\right)
{d\phi \over d\lambda}(\lambda ) \cr
&\qquad +{2\over D_{oq}}\int^{\lambda_q}_{\lambda_o}
\, d\lambda \left( \left( \lambda_o +\lambda_q\right)\lambda
-\lambda_o\lambda_q -\lambda^2\right)g^{(0)ik}{\partial \phi\over
\partial x^k\partial x^j}(\lambda ) \cr}\eqno{(47)}$$

\noindent
The details of our model lens then transform this into

$$\eqalign{ M^i{}_j&=\left( 1-2\phi_o\right) \delta^i_j \cr
&+{2\eta_o^2m\over a_o D_{oq} }\left[  I_{2,1}
-I_{0,3} +\left( \lambda_o+\lambda_q\right)I_{1,3}
-\lambda_o\lambda_qI_{2,3}\right] \delta^i_j \cr
&+{6\eta_o^2m\over a_o D_{oq} }\left[
I_{0,5}-\left( \lambda_o+\lambda_q\right)
I_{1,5}+\lambda_o\lambda_qI_{2,5} \right]x^i_lx^j_l \cr}\eqno{(48)}$$

\noindent
where we have put

$$I_{k,n}=\int^{\lambda_q}_{\lambda_o}\, d\lambda
{1\over \lambda^k\left[ \left( \lambda_l -\lambda\right)^2
+y_l^2+z_l^2 \right]^{n/2} }\eqno{(49)}$$

\noindent
As all of these integrals may be explicitly performed, a general
(but complicated and uninstructive)
expression for the magnification matrix may be written down.
For illustration, we
will consider the situation that the impact parameter is
much smaller than all the other physical sizes in the problem,
so that the integrals simplify to

$$\eqalign{ I_{2,5} &=-{4\over 3\lambda_l^2\theta^4D_{ol}^4}
-{5\over \lambda_l^6}\Delta \cr
I_{1,5} &=-{4\over 3\lambda_l \theta^4 D_{ol}^4}+{1\over \lambda_l^5}
\Delta \cr
I_{0,5} &= -{4\over 3\theta^4D_{ol}^4} \cr
I_{2,3} &=-{2\over \lambda_l^2\theta^2 D_{ol}^2} +{3\over \lambda_l^4}
\Delta \cr
I_{1,3} &=-{2\over \lambda_l\theta^2D_{ol}^2} +{1\over \lambda_l^3}
\Delta \cr
I_{0,3} &=-{2\over \theta^2 D_{ol}^2} \cr
I_{2,1} &=-{D_{lq}\over \lambda_l^2\lambda_q}+{D_{ol}\over
\lambda_l^2\lambda_o}+{1\over \lambda_l^2}\Delta \cr} \eqno{(50)}$$

\noindent
where

$$\Delta =\left[ -\sinh^{-1}\left(
{\lambda_l\left( \lambda_l -\lambda \right)
\over \lambda\theta D_{ol}}\right)
\right]_{\lambda_o}^{\lambda_q} \eqno{(51)}$$

With somewhat more algebra it can be shown that $I_{2,1}$ and
all the terms proportional to $\Delta$ contribute negligibly to
the magnification matrix under the assumption of small impact
parameter. The remaining terms combine to produce

$$M^i{}_j=-{4\eta_o^2mD_{lq}\over a_o\lambda_l^2D_{ol}D_{oq}\theta^2}
\delta^i_j +{8\eta_o^2mD_{lq}\theta^i\theta^j
\over a_o\lambda_l^2D_{ol}D_{oq}\theta^4}\eqno{(52)}$$

\noindent
which, because $q$ and $e$ are equivalent to the needed order, is
also expressible as

$$M^i_j=-{4m{\bar D}_{le}\over {\bar D}_{ol}{\bar D}_{oe}}\left(
{1\over \theta^2}\delta^i_j -2{\theta^i\theta^j \over \theta^4}\right)
\eqno{(53)}$$

\noindent
This agrees exactly with the usual magnification matrix for the
point mass lens (Schneider {\it et al.} 1993, Chapter 2).

\vskip0.5truein
\noindent
{\bf 6. Summary}

\nobreak
We have presented formulae (13), (14) for the null geodesics
intersecting an observer's worldline in an important class of
perturbed spacetimes, FRW backgrounds with scalar
perturbations, in the longitudinal gauge. We have used these equations
to obtain a general formula (36) for the magnification of ray bundles in
these spacetimes. With this, we can show
for the first time how the usual lens equation (1) and
magnification matrix are recovered in the curved FRW spacetimes without
dividing light paths into near and far lens regions. To illustrate
our formulae we have calculated the magnification matrix
appropriate to a point deflector in an Einstein-de Sitter
spacetime. We are able to show how the usual formula emerges along
with (in this case negligible) correction terms. The techniques
used in this paper are easily applicable to FRW spacetimes
with vector or tensor perturbations.

\vskip0.5truein
\noindent
{\bf Acknowledgements}

\nobreak
We thank Ramesh Narayan, Sylvanie Wallington, Uros Seljak,
and especially Sean Carroll for many helpful discussions.
This work was
supported by the National Science Foundation under grant AST90-05038.

\vfill\eject

\noindent
{\bf Appendix A: The Jacobi Equation of FRW Spacetimes}

We consider the equation of geodesic deviation in the spacetimes
$ds^{(0)2}$, related by conformal transformation to the background
FRW spacetimes. The equation of geodesic deviation along
$k^{(0)\mu}$ is

$${D^2\over d\lambda^2}\delta x^{\mu}=R^{(0)\mu}{}_{\alpha
\beta\gamma}k^{(0)\alpha}k^{(0)\beta}\delta x^{\gamma}. \eqno{({\rm A}1)}$$

\noindent
Here $D/d\lambda$ is the covariant derivative along $k^{(0)\mu}$.
We will take $k^{(0)\mu}$ to be the wavevector of a radial null geodesic
parameterized as in section 3 above. Changing variables via
$\delta x^{\mu}(\lambda )=P(\lambda ,a)^{\mu}{}_{\alpha}v^{\alpha}
(\lambda )$, $a$ an arbitrary affine parameter value along
$k^{(0)\mu}$, allows us to write (A1) as

$${d^2\over d\lambda^2}v^{\mu}(\lambda )=-\kappa J^{\mu}{}_{\alpha}
v^{\alpha}(\lambda) .\eqno{({\rm A}2)}$$

\noindent
We want to solve this equation subject to the initial
conditions $\delta x^{\mu}(u)=0$ and $d\delta x^{\mu}/d\lambda (u)=
\delta k^{\mu}(u)$. These initial conditions describe some
impulse which jolts the geodesic at some affine parameter distance
$u$.

(A2) conveniently decomposes into three separate
equations, one for each of the timelike component, and transverse
and longitudinal projections of the spatial components. To see
this, multiply (A2) by unity in the form of $\left(
\delta^{\alpha}_{\beta}-J^{\alpha}{}_{\beta} \right)+
J^{\alpha}{}_{\beta}$. The result is

$$\eqalign{ {d^2\over d\lambda^2} v^0 &= 0\cr
{d^2\over d\lambda^2}v^i_{\parallel} &=0 \cr
{d^2\over d\lambda^2}v^i_{\perp} &=-\kappa v^i_{\perp}\cr }
\eqno{({\rm A}3)}$$

\noindent
where $v^i_{\parallel}=\left(
\delta^i_{\beta}-J^i{}_{\beta} \right)v^{\beta}$ and
$v^i_{\perp}=J^i{}_{\beta}v^{\beta}$. The solutions
for the given boundary
conditions are elementary. Returning to the original variable
$\delta x^{\mu}$,

$$\eqalign{ \delta x^0(\lambda ) &={\gamma (\lambda ) \over \gamma (u)}
\delta k^0(u)(\lambda -u )\cr
\delta x^i_{\parallel}(\lambda ) &={\gamma (\lambda ) \over \gamma (u)}
\delta k^i_{\parallel} (\lambda -u )\cr
\delta x^i_{\perp}(\lambda ) &={\gamma (\lambda ) \over \gamma (u)}
\delta k^i_{\perp}\sin_{\kappa} (\lambda -u)\cr}
\eqno{({\rm A}4)}$$

We can use the solution above to determine the angular diameter
distance in the actual background spacetime.
To do this, let $k^{(0)\mu}(\lambda )=(1, -\gamma ,0,0)$ and
suppose an impulse at $u$ given by $\delta k^{\mu} (u)=\delta
k_{\perp}^{\mu}(u)=(0,0,-\gamma\epsilon ,0)$, with
$\epsilon$ infinitesimal. Let $\delta x^{\mu}(\lambda )$ be the
solution to the Jacobi equation for this impulse, $\delta x^{\mu}
(\lambda )=\gamma (\lambda )\sin_{\kappa } (\lambda -u)
(0,0,\epsilon ,0)=(0,0,r\epsilon ,0)$.
Since $\delta x^{\mu}$ is a Jacobi vector, $x^{(0)\mu}(\lambda )=
(\lambda , r,0,0)$ and $x^{(0)\mu}(\lambda )+\delta x^{\mu}(\lambda )$
are neighboring null geodesics to first
order in $\epsilon$. The angle that their wavevectors
make at $u$ is $\epsilon$. The proper linear distance
(in $d{\bar s}^{(0)2}$) that they
span at affine parameter $\lambda$ is $\left( \delta x^{\mu}
{\bar g}^{(0)}_{\mu\nu} \delta x^{\nu}\right)^{1/2}$.
The angular-diameter distance of our FRW background
is defined as the ratio of this
proper linear distance to the subtended angle,

$$\eqalign{ {\bar D}(u ,\lambda )&={1\over \epsilon}\left( \delta x^{\mu}
{\bar g}^{(0)}_{\mu\nu} \delta x^{\nu}\right)^{1/2} \cr
 &=a(\lambda )\sin_{\kappa}(u-\lambda )\cr} \eqno{({\rm A}5)}$$

\noindent
where the sense of the affine parameter
in (A5) is that $\lambda \le u$.

\vfill\eject

\noindent
{\bf Appendix B: Geodesic Deviation in Perturbed Spacetimes}

Consider an arbitrary metric perturbed spacetime with metric

$$g_{\mu\nu}=g^{(0)}_{\mu\nu}+h_{\mu\nu} \eqno{({\rm B}1)}$$

\noindent
with $h_{\mu\nu}$ small.  Let $x^{(0)\mu}_A(\lambda )$
be an affinely parametrized geodesic of $g^{(0)}_{\mu\nu}$.
Solutions, $\zeta^{(0)\mu}( \lambda )$,
to the Jacobi equation of the background along $x^{(0)\mu}_A(\lambda )$,

$$ {d^2\zeta^{(0)\mu}\over d\lambda^2}+2\Gamma^{(0)\mu}_{\alpha\beta}
k^{(0)\alpha}{d\zeta^{(0)\beta}\over d\lambda }+\Gamma^{(0)\mu}_{\alpha
\beta ,\sigma}k^{(0)\alpha}k^{(0)\beta}\zeta^{(0)\sigma}=0
\eqno{({\rm B}2)}$$

\noindent
generate nearby geodesics, $x^{(0)\mu}_B(\lambda )$,
via $x^{(0)\mu}_B(\lambda )=x^{(0)\mu}_A
(\lambda )+\zeta^{(0)\mu}(\lambda )$. Let $x^{\mu}_A(\lambda )$ and
$x^{\mu}_B(\lambda )$ be geodesics of $g_{\mu\nu}$ generated
using the perturbative geodesic expansion from $x^{(0)\mu}_A(\lambda )$
and $x^{(0)\mu}_B(\lambda )$ respectively. We claim that, to
first order, $x^{\mu}_B(\lambda )-x^{\mu}_A(\lambda )$ solves
the geodesic deviation equation of $g_{\mu\nu}$ along
$x^{\mu}_A(\lambda )$.

There is certainly nothing surprising in this claim. In fact, the
assertion must be true if the PGE correctly generates nearby
geodesics of $g_{\mu\nu}$ as we claim it does. The explicit proof
offered here can, thus, be thought of as another check on the method
itself. Since the proof is nothing more than a tedious application
of the usual perturbative techniques we present it only in schematic
form.

To see that our assertion is true,
we start with the Jacobi equation of $g_{\mu\nu}$ along
$x^{\mu}_A(\lambda )$

$$ {d^2w^{\mu}\over d\lambda^2}+2\Gamma^{\mu}_{\alpha\beta}
k_A^{\alpha}{dw^{\beta}\over d\lambda }+\Gamma^{\mu}_{\alpha
\beta ,\sigma}k_A^{\alpha}k_A^{\beta}w^{\sigma}=0
\eqno{({\rm B}3)}$$

\noindent
By a Taylor expansion, the decomposition of the Christoffel terms into
zeroth and first order expressions in the perturbation, the
ansatz $w^{\mu}=w^{(0)\mu}+w^{(1)\mu}$,
and the PGE identity $k_A^{\mu}=k_A^{(0)\mu}+k^{(1)\mu}$, we can
express equation (B3) as two equations holding along $x^{(0)\mu}_A
(\lambda )$, one containing only zeroth order terms and the other
containing only first order terms. The zeroth order equation is
seen to be equivalent to (B2), so that we already know that
$w^{(0)\mu}(\lambda )=x^{(0)\mu}_B(\lambda )-x^{(0)\mu}_A(\lambda )$
is a solution.
The first order equation may be written $T\left( w^{(1)}\right)=0$
for some operator $T$. The specific form of $T$ will not be necessary
for our current purposes but it is not hard to obtain.

Consider now the equation obeyed by $x^{(1)\mu}_B(\lambda )=
x^{\mu}_B(\lambda )-x^{(0)\mu}_B(\lambda )$,

$${d^2x^{(1)\mu}_B\over d\lambda^2}+2\Gamma^{(0)\mu}_{\alpha\beta}
k_B^{(0)\alpha}{dx^{(1)\beta}_B\over d\lambda }+\Gamma^{(0)\mu}_{\alpha
\beta ,\sigma}k_B^{(0)\alpha}k_B^{(0)\beta}x^{(1)\sigma}_B=
\Gamma^{(1)\mu}_{\alpha\beta}k_B^{(0)\alpha}k_B^{(0)\beta}
\eqno{({\rm B}4)}$$

\noindent
which holds along $x^{(0)\mu}_B(\lambda )$ (PB). Using Taylor and
$ k^{(0)\mu}_B(\lambda )=k^{(0)\mu}_A(\lambda )+{\dot w}^{(0)\mu}
(\lambda )$, where $\cdot\equiv d/d\lambda $, we can write
(B4) as an equation along $x^{(0)\mu}_A(\lambda )$. If we subtract
from this equation the equation for $x^{(1)\mu}_A(\lambda )$, which
also holds along $x^{(0)\mu}_A$ (and is identical to (B4) after the
subscript $B$'s are replaced by $A$'s) we obtain the result
$T\left( x^{(1)}_B -x^{(1)}_A\right)=0$ and our assertion is proved.

\vfill\eject

\noindent
{\bf References}

\vskip 12pt
\normalbaselineskip=8pt plus0pt minus0pt
                            \parskip 0pt

\def\ref#1  {\noindent \hangindent=24.0pt \hangafter=1 {#1} \par}
\def\vol#1  {{\bf {#1}{\rm,}\ }}
\ref{De Felice, F., \& Clarke, C. J. S. 1990, Relativity on Curved
Manifolds (Cambridge: Cambridge University Press)}
\ref{Dyer, C. C., \& Roeder, R. C. 1972, ApJ,
174, L115}
\ref{Dyer, C. C., \& Roeder, R. C. 1973, ApJ,
180, L31}
\ref{Ehlers, J., \& Schneider, P. 1986, A\&A,
168, 57}
\ref{Ellis, G. F. R. 1971, in Proceedings of the International School of
Physics, ``Enrico Fermi," General Relativity and Cosmology,
ed. R. K. Sachs (New York: Academic Press), 104}
\ref{Futamase, T. 1989, MNRAS, 237, 187}
\ref{Futamase, T., \& Sasaki, M. 1989, Phys. Rev.
D., 40, 2502}
\ref{Jacobs, M. W., Linder, E. V., \& Wagoner, R.~V. 1992,
Phys. Rev. D., 45, R3292}
\ref{Kaiser, N. 1992, ApJ, 388, 272}
\ref{Kamionkowski, M., Ratra, B., Spergel, D. N., \& Sugiyama, N.
1994, ApJ, 434, L1}
\ref{Kamionkowski, M., Spergel, D. N., \& Sugiyama, N.
1993, ApJ, 426, L57}
\ref{Kristian, J., \& Sachs, R. K. 1965, ApJ, 143, 379}
\ref{McVittie, G. C. 1964, General Relativity and Cosmology
(2d ed., London: Chapman and Hall}
\ref{Pyne, T. \& Birkinshaw, M. 1993, ApJ, 415, 459}
\ref{Refsdal, S. 1964, MNRAS, 128, 295}
\ref{Sachs, R. K. 1961, Proc. Roy. Soc. Lond., A264, 309}
\ref{Sasaki, M. 1993, Prog. Theor. Phys., 90, 753}
\ref{Schneider, P., Ehlers, J., \& Falco, E.~E. 1993,
Gravitational Lensing (Berlin: Springer-Verlag)}
\ref{Seitz, S., Schneider, P., \& Ehlers, J. 1994, Class. \& Q. Grav.,
11, 2345}
\ref{Seljak, U. 1994, ApJ, 436, 509}
\ref{Spergel, D. N., Pen, U. L., Kamionkowski, M., \& Sugiyama, N.
1993, to
appear in Proceedings of the Nishonomaya Yukawa Memorial Symposium,
Nishonomaya, Japan, 29 Oct. 1993}
\ref{Synge, J. L. 1960, Relativity: The General Theory
(Amsterdam: North-Holland)}
\ref{Watanabe, K., Sasaki, M., \& Tomita, K. 1992,
ApJ, 394, 38}
\ref{Watanabe, K., \& Tomita, K. 1990, ApJ, 355, 1}

\normalbaselineskip=24pt plus0pt minus0pt
                  \parskip 12.0pt
\vfill\eject

\noindent
{\bf Figure 1.} Gravitational lensing by a single symmetric lens.
$\beta$ is the angle at the observer, $o$, between the image of the
lens, $l$, and the unlensed image of the emitter, $e$. $\theta$ is
the angle between the image of the lens and the light ray $epo$,
along which the emitter is observed. The point of deflection, $p$,
lies in the lens plane.
The intersection of the line joining the observer to
the point of deflection with the source
plane is labeled $p^{\prime}$. $\alpha$, the
deflection angle, is the angle at $p$ between the image of the emitter,
$e$, and $p^{\prime}$. $D(o,l)$ is the distance between the observer
and the point of deflection, $D(o,e)$ is the distance between the
observer and the source plane, $D(l,e)$ is the distance between
the lens and source planes.

\noindent
{\bf Figure 2.} The lensing of a congruence of null geodesics.
The actual congruence, with central ray $x^{\mu}$, joins the observer
at $o$ with the emitter at $e$ and has area $dA$ at the emitter.
It is constructed from a congruence of the background, with
central geodesic $x^{(0)\mu}$, reaching between the observer and
a point $q$. The redshift in the background between the observer
and $q$ is equal to the redshift in the perturbed spacetime
between the observer and the emitter. The background congruence
has area $dA^{(0)}$ at $q$.

\noindent
{\bf Figure 3.} The geometry of Section 4. $\theta$ is the angle at
the observer between the lens and the emitter. $\lambda_l$ is the affine
parameter value at which the background geodesic intersects the
lens plane. $\lambda_e$ is the affine
parameter value at which the background geodesic intersects the
source plane. The observer is located at affine parameter value
$\lambda_o$. $D_{ol}$ is the angular diameter distance
in the background between the
observer and the lens plane.

\vfill
\eject

\end